\documentclass[condensedmatter,preprints,article,accept,moreauthors,pdflatex]{Definitions/mdpi}

\usepackage{xcolor}
\usepackage{dsfont}
\usepackage{amsmath}
\usepackage{amssymb}
\usepackage{booktabs}
\usepackage{adjustbox}

\firstpage{1} 
\pubvolume{xx}
\issuenum{1}
\articlenumber{5}
\pubyear{2020}
\copyrightyear{2020}
\history{Received: date; Accepted: date; Published: date}

\Title{Transforming Lindblad Equations into  Systems of Real-Valued Linear Equations: Performance Optimization and Parallelization of an Algorithm}

\Author{Iosif Meyerov $^{1}$, Evgeny Kozinov $^{1}$, Alexey Liniov$^{1}$, Valentin Volokitin$^{1}$, Igor Yusipov$^{1,2}$, Mikhail Ivanchenko$^{1,2}$,
and Sergey Denisov $^{3,1}$*}

\AuthorNames{Iosif Meyerov, Evgeny Kozinov, Alexey Liniov, Valentin Volokitin, Igor Yusipov, Mikhail Ivanchenko,
and Sergey Denisov}

\address{%
$^{1}$ \quad Mathematical Center, Lobachevsky University, 603950 Nizhni Novgorod, Russia\\
$^{2}$ \quad Department of Applied Mathematics, Lobachevsky University, 603950 Nizhni Novgorod, Russia\\
$^{3}$ \quad Department of Computer Science, Oslo Metropolitan University, N-0130 Oslo, Norway}

\corres{Correspondence: sergiyde@oslomet.no}

\abstract{
With their constantly increasing peak performance and memory capacity, modern supercomputers offer new 
perspectives on numerical studies of open many-body quantum systems. These systems are often modeled
by using Markovian quantum master equations describing the evolution of the system density operators. 
In this paper we address master equations of the Lindblad form,  
which are a popular theoretical tool in quantum optics, cavity quantum electrodynamics, and optomechanics. By using the 
generalized Gell-Mann matrices as a basis, any Lindblad equation can be transformed into a system of ordinary differential 
equations with real coefficients. 
Recently we presented an implementation  of the transformation
with the computational complexity 
scaling as $O(N^5 log N)$ for dense Lindbaldians and $O(N^3 log N)$ for sparse ones. However, infeasible 
memory costs remains a serious obstacle on the way to large models. Here we present a parallel cluster-based implementation of the algorithm
and demonstrate that it allows us to integrate a sparse Lindbladian  model of the dimension $N=2000$ 
and a dense random Lindbladian model of the dimension $N=200$ by using $25$ nodes with 64 GB RAM per node.}

\keyword{open quantum systems, Lindblad equation, parallel computing, MPI, performance optimization}

\begin{document}


\section{Introduction}
High-performance computation technologies 
are becoming more and more important for modeling of complex quantum systems, 
both as a tool of theoretical research \cite{c1a,c1b} and a mean to explore possible technological applications \cite{c2a,c2b}. 
From the computational point of view, to simulate a coherent $N$-state quantum system, i.e., a system  that is completely isolated 
from its environment, we have to deal with a generator of evolution in the form of an  $N \times N$ 
Hermitian matrix. When modeling an \textit{open} system \cite{c3}
that is described with its density operator, we have to deal with \textit{superoperators}, generators of the dissipative evolution 
represented by $N^2 \times N^2$ matrices. Evidently, numerical simulations of open systems require  large memory  and  longer
computation time than in the case of  coherent models of equal size. This is a strong motivation for the development of new 
algorithms and implementations that can utilize possibilities provided by modern supercomputers.

In this paper, we consider generators of open quantum evolution of the so-called
Gorini–Kossakowski–Sudarshan–Lindblad (GKSL) form \cite{c3a,c3b,c3} (henceforth also called ‘Lindbladian’).
The corresponding master equations are popular tools to model dynamics of open systems  in quantum optics \cite{cC}, 
cavity quantum electrodynamics \cite{cD}, and quantum optomechanics \cite{cE}. 
More precisely, we address an approach which transforms an arbitrary GSKL equation into a system of linear 
ordinary differential equations (ODEs) with real coefficients \cite{c11}.

The idea of such transformation is well known since the birth of the GSKL equation \cite{c3a,chur}. 
For a single qubit, it corresponds to the Bloch-vector representation and results in the Bloch equations \cite{c11}. 
For an $N=3$ system, it can be realized with eight Gell-Mann matrices \cite{gel}.  For any $N > 3$ it can be performed  
by using a complete set of infinitesimal generators of the $SU(N)$ group \cite{Li}, rendering density matrix in form of 
`coherence-vector'  \cite{c11}. 
We are not aware of any implementation of this procedure for $N > 4$; as we discuss below, the complexity of the procedure grows quickly with $N$.

When implementing the expansion over the $SU(N)$ generators in our prior work \cite{c12}, 
we were mainly driven by the interest to technical aspects of the implementation.
We also thought that the expansion could be of interest in the context of possible use of 
existing highly-efficient  numerical methods to integrate large ODE systems \cite{parallel}. 
Very recently it turned out that the expansion itself plays a key role in a notion  of an ensemble of random Lindbladians \cite{randomL}, a generalazitaion of the idea of  
the GUE ensemble (that is a totally random Hamiltonian) \cite{edelman} to GSKL generators (we  sketch the definition in Section III). 
In this respect, it is necessary to go for large $N$ in order to capture universal asymptotic properties (including scaling relations).
The upper limit reported in Ref.~\cite{randomL} is $N=100$.

The implementation proposed in Ref.~ \cite{c12} includes two main 
steps that are (i) Data Preparation (calculation of expansion coefficients, which form the ODE system) and (ii) Integration of 
the obtained ODE system. Step (i) is complexity dominating; its complexity scales as $N^5 log N$ for dense Lindbladians and $N^3 logN$ 
for sparse ones \cite{c12}. The implementation of the algorithm posed another problem that is infeasible memory 
requirements to solve large models. Due to the complexity of the algorithm, the solution of this problem is not 
straightforward. Here we propose a parallel version of the algorithm that distributes memory costs across several cluster 
nodes thus allowing for an increase of the model size up to $N=2000$   
and up to $N=200$, in the sparse and dense (random Lindbladians) cases, respectively.



\section{Expansion over the basis of $SU(N)$ group generators}
\label{sec:2}

We consider a GKSL equation \cite{c3a}, $\dot{\rho}_t = \mathcal{L}(\rho_t)$, with the Lindbladian of the following general form:
\begin{eqnarray}  \label{L0}
\mathcal{L}(\rho) = -i[H(t),\rho]+\mathcal{L}_D(\rho), 
\label{eq:0}
\end{eqnarray}
with time-dependent (in general) Hamiltonian $H(t)$ and a dissipative part 
\begin{eqnarray}  \label{L1}
\!\!\! \mathcal{L}_D(\rho) = 
\!\! \!\sum\limits_{m,n=1}^{N^2-1} \!\! K_{mn} \Big(F_n\rho F^{\dagger}_m - \frac 12 (F^{\dagger}_m F_n\rho+\rho F^{\dagger}_m F_n)\Big),~~
\label{eq:1}
\end{eqnarray}
where Hermitian matrices  $\{F_n\}$, $n = 1,2,3,\ldots, M = N^2-1$ form a set of infinitesimal generators of $SU(N)$ group \cite{c3a}.
They satisfy orthonormality condition, $\mathrm{Tr}(F_n F_m^\dagger) = \delta_{n,m}$, and provide, together with the identity 
operator, $F_0 = \mathds{1}/N$, a  basis to span a Hilbert-Schmidt space of the dimension $N$.  
The {\em Kossakowski matrix}  $K = \{K_{mn}\}$  is positive semi-definite and 
encodes the action of the environment on the system.

Properties of the set  $\{F_n\}$ are discussed Ref.~\cite{c11}; 
here we only introduce definitions, equations and formulas necessary for describe the algorithm.
The matrix set consist of matrices of the three types, $\{F_i \}=(\{S^{(j,k)} \},\{J^{(j,k)} \},\{D^l \})$, where

\begin{eqnarray}  \label{eq_sf}
S^{(j,k)}= \frac{1}{\sqrt{2}} (e_j e_k^T+e_k e_j^T ), 1 \leq j < k  \leq N,
\end{eqnarray}

\begin{eqnarray}  \label{eq_jf}
J^{(j,k)} = \frac{-i}{\sqrt{2}} (e_j e_k^T-e_k e_j^T ), 1  \leq j < k  \leq N,
\end{eqnarray}

\begin{eqnarray}  \label{eq_df}
D^l=\frac{i}{\sqrt{l(l+1)}} (\sum_{k=1}^l{(e_k e_k^T ) - e_{l+1} e_{l+1}^T}), 1 \leq l \leq N-1.
\end{eqnarray}

Any Hermitian matrix can be expanded over the basis $\{F_i\}$, 
\begin{equation}
	A=a_0 F_0 + \sum_{j=1}^M{a_j F_j}, ~~a_0=\frac{Tr(A)}{N},~~a_j=Tr(F_j A),~a_j \in R.
	\label{eq:02}
\end{equation}

The expansion of the density operator 
\begin{equation}
\rho(t) = F_0 + \sum_{i=1}^M{v_i(t) F_i}
\label{eq:rho}
\end{equation}
yields the ''Bloch vector'' \cite{c11,c17}: 
$\overrightarrow{v}=(v_1,...,v_M) \in R^M$. The Kossakowski matrix can be diagonalized, $\tilde{K} = SKS^\dag = 
\mathrm{diag}\{\gamma_1, \gamma_2,..., \gamma_P\}$, $P \leq M$. 
By using spectral decomposition, $K = \sum_{p=1}^{P} \gamma_p {\bar{l}_p\bar{l}_p^\dag}$, the dissipative part of the Lindbladian, Eq.~(\ref{eq:1}),
can be recast into
\begin{eqnarray}
\mathcal{L}_D (\rho) = \frac{1}{2}\sum_{p=1}^{P} \gamma_p  \sum_{j,k=1}^{M} {l_{p;j} l^{\ast}_{p;k}\left([F_j, \rho F_k^\dag ]+[F_j \rho,F_k^\dag ] \right)}.	
	\label{diss_new}
\end{eqnarray}
Now we can transform the original GKSL equation into 
\begin{equation}
\sum_{i=1}^M{\frac{dv_i}{dt} F_i} = -i \sum_{i,j=1}^M{h_j(t) v_i [F_j,F_i ]} + \frac{1}{2}\sum_{p=1}^{P} \gamma_p \sum_{i,j,k=1}^M{l_j \overline{l_k} v_i ([F_j,F_i F_k^\dagger]+[F_j F_i,F_k^\dagger ])}.
	\label{eq:03}
\end{equation}
Here $\{h_j(t)\}$ are coefficients of the Hamiltonian expansion, $H(t) =  \sum_{i,j=1}^M h_j(t)F_j$.

This can be represented as a non-homogeneous ODE system,
\begin{equation}
	\frac{dv(t)}{dt}=(Q(t)+R)v(t)+K,
	\label{eq:04}
\end{equation}
Matrices $Q(t)$, $R$, and the vector $K$ are calculated by using the following expressions~\cite{c11,c12} (their entries are denoted with 
lower-case versions of the corresponding symbols):

\begin{equation}
	f_{mns}=-i\mathrm{Tr}(F_s [F_m,F_n ]),~d_{mns} = \mathrm{Tr}(F_s {F_m,F_n }),~m,n,s=1,\dots,M
	\label{eq:05}
\end{equation}

\begin{equation}
	z_{mns}=f_{mns}+i d_{mns},~m,n,s=1,\dots,M
	\label{eq:06}
\end{equation}

\begin{equation}
	q_{sn}= \sum_{m=1}^M{h_m f_{mns}},~s,n=1,\dots,M
	\label{eq:07}
\end{equation}

\begin{equation}
	k_s=\frac{i}{N} \sum_{m,n=1}^M{l_m \overline{l_n} f_{mns}},~s=1,\dots,M
	\label{eq:08}
\end{equation}

\begin{equation}
	r_{sn}= -\frac{1}{2}\sum_{p=1}^{P} \gamma_p \sum_{j,k,l=1}^M{l_j \overline{l_k}(z_{jln} f_{kls}+\overline{z_{kln}}f_{jls} )},~s,n=1,\dots,M
	\label{eq:09}
\end{equation}
Note that we consider the case of time-independent dissipation [see the r.h.s. of Eq.~(1)], matrix $R$ and vector $K$ in Eq.~(\ref{eq:04}) 
are time independent. The vector appears as a result of the Hermiticity of Kossakowski matrix $K$ \cite{c11}. Finally, 
Bloch vector $\overrightarrow{v}(t)$  can be  easily converted back 
into the density operator $\rho(t)$.

\section{Models}
\label{sec:3}

We consider two test-bed models, with a sparse and dense Lindbladians.

The first model describes $N-1$ indistinguishable interacting bosons, which are hopping between the sites of a periodically rocked dimer . 
The model is described with a time-periodic Hamiltonian \cite{weiss, vardi, trimborn, poletti, tomkovic}, 
\begin{equation}
	H(t)=-J(b_1^\dagger b_2+b_1 b_2^\dagger )+ \frac{2U}{N-1} \sum_{j=1}^2{n_j (n_j-1)}+\varepsilon(t)(n_2-n_1 ),
	\label{eq:10}
\end{equation}
where $b_j$ and $b_j^\dagger$ are the annihilation and creation operators of an atom at site $j$, while $n_j=b_j^\dagger b_j$ is 
the operator of number of particle on $j$-th site, $J$ is the tunneling amplitude, $\frac{2U}{N-1}$ is 
the interaction strength (normalized by a number of bosons), and $\varepsilon(t)$ represents the modulation of 
the local potential. $\varepsilon(t)$ is chosen as $\varepsilon(t)=\varepsilon(t+T)=E+A\Theta(t)$, where $E$ is 
the stationary energy offset between the sites, and $A$ is the dynamic offset. The modulating function is defined as 
$\Theta(t) = 1$ for $0 \leq t < T / 2$, $\Theta(t) = -1$ for $T / 2 \leq t < T$

The dissipation part of the Lindbladian has the following form \cite{DiehlZoller2008}:
\begin{equation}
	\mathcal{L}_D(\rho) =\frac{\gamma}{N-1}\Big(L\rho L^{\dagger} - \frac 12 (L^{\dagger} L\rho+\rho L^{\dagger} L)\Big),~~ L = (b_1^\dagger+b_2^\dagger)(b_2+b_1 ).
	\label{eq:11}
\end{equation}
This dissipative coupling tries to ''synchronize'' the dynamics on the sites by constantly recycling 
antisymmetric out-phase mode into symmetric in-phase one. The non-Hermiticity of $L$ guarantees that the asymptotic state $\rho_{\infty}$, 
$\mathcal{L}(\rho_{\infty}) = 0$, is different from the normalized identity $\mathds{1}/N$ (also called 'infinite temperature state').
Both matrices, of  Hamiltonian $H$ and  of Kossakowski matrix $K$, are sparse [$P=1$ in Eq.~(5)].  
Correspondingly, both matrices on the rhs of Eq.~(\ref{eq:04}), $Q(t)$ and $R$, are sparse.

Numerical experiments were performed with 
the following parameter values: $J=-1.0$, $U=2.2$, $E=-1.0$, $A=-1.5$, $\Theta(t)=\mathrm{sign}(t-\pi)$, $T=2\pi$, $\gamma=0.1$, $N=100,\dots,2000$.

The second model is a random Lindbladian model recently introduced in Ref.~\cite{randomL}.
It has $H \equiv 0$ and the dissipative part of the Lindbladian, Eq.~(\ref{eq:1}), is
specified by a random Kossakowski matrix. Namely, it is generated from an ensemble of complex Wishart matrices \cite{c18},
$W = GG^\dagger\ge 0$, where $G$ is a complex  $N^2-1 \times N^2-1$ Ginibre matrix. We use the normalization condition 
$\mathrm{Tr}K = N$, that is, $K = N GG^\dagger/\mathrm{Tr}GG^\dagger$. By construction, the Kossakowski matrix is fully dense. 
Therefore, matrix $R$ on the rhs of Eq.~(\ref{eq:04}) is dense.

To obtain the corresponding Lindbladian, we need to 'wrap' the Kossakowski matrix into basis states, $\{F_n\}$, according to Eq.~(\ref{eq:1}).
Following the nomenclature introduced in Section I, this corresponds to step (i). 
If we want to explore spectral features of the Lindbladian, the actual propagation, step (ii), is not needed. However, 
it could be needed in some other context so we will adress it also.

A generalization of the random Lindbladian model to many-body setup was proposed very recently \cite{david}.
It allows  to take into account the topological range of interaction in a 
spin chain model, by controlling number of neighbors $n$ involved into an $n$-body Lindblad operator $\mathcal{L}^i_{\{n\}}$ acting on 
the $i$-th spin. The total Lindbladian is therefore  $\mathcal{L}_D = \sum_i \mathcal{L}^i_{\{n\}}$.

\begin{table}[ht]
	\centering
	\caption{The algorithm}
	\label{tab:01}
	\begin{adjustbox}{max width=0.95\textwidth}
	\begin{tabular}{@{}lll@{}}
		\toprule
		\multicolumn{1}{c}{\textbf{Step}} & \multicolumn{1}{c}{\textbf{Substep (the sequential algorithm)}}          & \multicolumn{1}{c}{\textbf{Parallelization}}    \\ \midrule
		\textbf{1. Initialization}          & \begin{tabular}[t]{@{}l@{}}1.1. Read the initial data \\from configuration files.\\ 1.2. Allocate and initialize memory.\end{tabular}                                                                                                                                                                                         & \begin{tabular}[t]{@{}l@{}}Sequential step, all the operations are \\ performed on every node of a cluster.\end{tabular}                                                                                                                                                                              \\ \midrule
		
		                                    & \textbf{Data Preparation cycle:} & \begin{tabular}[t]{@{}l@{}}This step is parallelized, computation time \\ and memory costs are distributed among \\ cluster nodes via Message Passing Interface (MPI).\end{tabular}\\
    & &\\                                        
																	& \begin{tabular}[t]{@{}l@{}}  2.1. Compute the coefficients $h_i$, $l_i$ \\ of the expansion of the matrices $H$ and $L$ \\ in the basis $\{F_i\}$. \end{tabular}& \begin{tabular}[t]{@{}l@{}} Step 2.1. (Figure~\ref{fig:image1}, Panel A) is not resource demanding \\ and, therefore, is performed on each cluster node independently. \end{tabular}\\											
& & \\
		\begin{tabular}[t]{@{}l@{}}\textbf{2. Data }\\ \textbf{Preparation}\end{tabular}&
  		                                    \begin{tabular}[t]{@{}l@{}} 2.2. Compute the coefficients \\ $f_{ijk}$, $d_{ijk}$, $z_{ijk}$ by formulas (\ref{eq:05},\ref{eq:06}). \end{tabular}& \begin{tabular}[t]{@{}l@{}} Step 2.2. (Figure~\ref{fig:image1}, Panel B) is memory demanding\\ in a straightforward implementation. Unlike the \\ sequential algorithm, we compute only nonzero\\ elements of the tensors on the fly,  when they are needed. \end{tabular} \\
																					& &\\
											& \begin{tabular}[t]{@{}l@{}} 2.3. Compute the coefficients $q_{sm}$ \\ by formula (\ref{eq:07}).\\ 2.4. Compute the coefficients $k_s$\\ by formula (\ref{eq:08}).\\ 2.5. Compute the coefficients $r_{sm}$ \\ by formula (\ref{eq:09}). \end{tabular} &  \begin{tabular}[t]{@{}l@{}}Parallel steps 2.3, 2.4, and 2.5 are sketched\\ in Figure~\ref{fig:image1} (Panels C, D, and E, respectively). \\ These steps are time and memory consuming \\ and are executed in parallel on cluster nodes. \end{tabular}\\
											& & \\
											& \begin{tabular}[t]{@{}l@{}} 2.6. Compute the initial value $v(0)$. \end{tabular} &  \begin{tabular}[t]{@{}l@{}} Step 2.6 is not resource demanding and is realized\\ on each cluster node independently. \end{tabular}\\ \midrule
			 \begin{tabular}[t]{@{}l@{}}\textbf{3. ODE}\\ \textbf{Integration}\end{tabular}     & \begin{tabular}[t]{@{}l@{}}3.1. Integrate the ODE (\ref{eq:04}), \\over time to $t = T$ \\ by means of the Runge-Kutta method.\\ 3.2. Compute $\rho(T)$ by formula (\ref{eq:rho}).\end{tabular}                                                                                                                                                                                                                                                                                                                                                                                     & \begin{tabular}[t]{@{}l@{}}This step is parallelized via MPI (among \\ cluster nodes), OpenMP (among CPU cores\\ on every node), and SIMD instructions \\ (on every CPU core).\end{tabular}  \\ \midrule
\textbf{4. Finalization}            & \begin{tabular}[t]{@{}l@{}}4.1. Save the results.\\ 4.2. Release memory.\end{tabular}                                                                                                                                                                                                                                          & \begin{tabular}[t]{@{}l@{}}Here we gather results from computational\\ nodes, save the results, and finalize MPI.\end{tabular}                                                                                                                                                                             \\ \bottomrule
	\end{tabular}
	\end{adjustbox}
\end{table}

\section{Algorithm}
\label{sec:4}

\begin{figure}[H]
	\centering
	\includegraphics{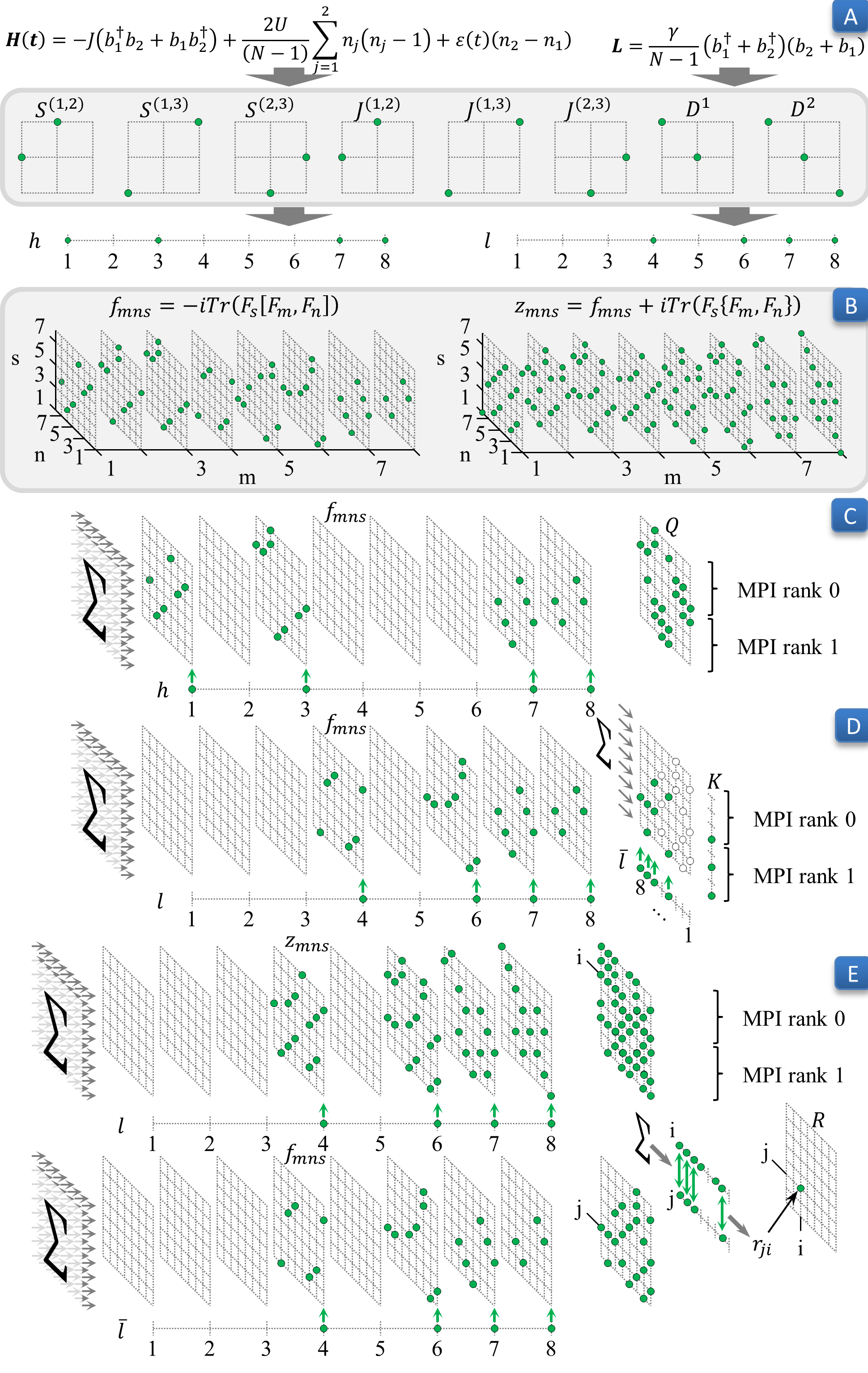}
	\caption{The parallel data preparation pipeline for the dimer model. 
	On Panel A we sketch distribution of nonzero elements of matrices  $S$, $J$, and $D$, forming basis $\{F\}$, Eqs.~(\ref{eq_sf}), (\ref{eq_jf}), and (\ref{eq_df}), respectively. 
	Panel B depcits locations of nonzero elements in tensors $f$ and $d$, Eq.~(\ref{eq:05}), 
	which are not stored in memory but computed on the fly, during the Data Preparation step. 
	Panels C, D, and E show how matrices Q, Eq.~(\ref{eq:07}), K, Eq.~(\ref{eq:08}), and R, Eq.~(\ref{eq:09}), are computed in parallel by two MPI processes (see Table I and Section~4, chapters 2.3 -2.5, 
	for more details).}
	\label{fig:image1}
\end{figure}

In our recent paper \cite{c12}, we presented a detailed description of the sparse and 
dense data structures and sequential algorithm to perform the above discussed expansion and the subsequent propagation of the ODE system. 
Below we present a pseudocode of the previous sequential and new parallel algorithms (Table~\ref{tab:01}) and briefly overview both algorithms. A particular realization of the memory demanding and complicated part of the parallel algorithm
for the dimer with two bosons is sketched in Figure~\ref{fig:image1}.

\textbf{1. Initialization (sequential; performed on every node of a cluster).} We load initial data from configuration files, allocate and initialize 
necessary data structures and perform some pre-computing. It takes $O(N^2)$ of time and 
$O(N^2)$of  memory only and, therefore, can be done sequentially on every computational node.

\textbf{2. Data Preparation and its parallelized via Message Passing Interface (MPI).} 
During this step only nonzero values of data structures are calculated. 
This improves the software performance by several orders of magnitude as compared to a naive implementation \cite{c12}. 
This step requires $O(N^5 log N)$ operations and $O(N^4)$ memory for dense Lindbladians
and $O(N^3 log N)$ operations and $O(N^3)$ memory for sparse ones \cite{c12}.

Unfortunately, this approach leads to infeasible memory requirements in a sequential 
mode when exploring models of large sizes. Thus, on a single node with 64 GB RAM 
we can study models with dense matrices of size up to $N \backsimeq 100$ and sparse matrices $H$ and $L$ of size $N=1000$.

Below we briefly explain which stages are the most time and memory consuming 
and the origin of the asymptotic complexity scalings. We also show how to overcome the 
infeasible memory requirements by using a parallel data preparation algorithm which 
distributes data and operations among cluster nodes.

\textbf{2.1. Compute the expansion coefficients $h_j$, $l_j$ for matrices $H$ and $L$.} 
Each element of the vectors $h_j$ and $l_j$ corresponds to the product of one of the matrices $\{F_i\}$ and 
matrices $H$ and $L$, respectively. Based on the specific sparse structure of the matrices $\{F_i\}$, the corresponding
coefficients can be computed in one pass over nonzero elements of the $H$ and $L$ matrices 
that requires $O(N^2)$ operations and produces vectors $h$ and $l$ with $O(N^2)$ nonzero elements in case of dense matrices 
and $O(N)$ for sparse ones.  This step is performed by each MPI-process independently.

\textbf{2.2. Compute the coefficients $f_{mns}$, $d_{mns}$, $z_{mns}$ by formulas (\ref{eq:05})-(\ref{eq:06}).} 
Due to the sparsity structure of the matrices $\{F_i\}$, most of the coefficients $f_{mns}$, $d_{mns}$, $z_{mns}$ 
are equal to zero. Only $O(N^3)$ of them have nonzero values, where each coefficient can be computed in $O(1)$ time.
In this algorithm, nonzero coefficients of the tensors are calculated at the moment when they are required 
in the calculations (steps 2.3-2.5 of the algorithm). Therefore all three tensors are not stored in memory.

\textbf{2.3. Compute the coefficients $q_{sm}$ by formula (\ref{eq:07}).} During this and two next steps, 
distribution of operations with the tensor $F$ among MPI processes by the index $s$ is performed. 
Each process calculates a set of nonzero elements $h_n \times f_{nms}$ and the 
corresponding panel of the matrix $Q$. Then all the panels are collected into resulting matrix by the master process. 
In case of a dense matrix $H$, time and memory complexity of this step is proportional to the number of $f_{mns}$ coefficients which is $O(N^3)$ and it is much less for sparse $H$.
Note that a uniform distribution of ranges of index values $s$ between MPI processes can lead 
to a large imbalance in a number of operations and memory requirements. To overcome this problem we employ a two-stage load balancing scheme. First, 
we compute the number of non-zero entries in the rows of resulting matrix. 
Next, we distribute rows between cluster nodes providing approximately the same number of elements on every node.

\textbf{2.4. Compute the coefficients $k_s$ by formula (\ref{eq:08}).} Calculation of the vector $K$ uses the 
same balanced distribution of operations with tensor $F$ between MPI processes. Each process computes non-zero terms $l_n \times f_{nms}$ and calculates a block of vector $K$. 
All MPI processes send results to the master process, which assembles them into a single vector.
For dense matrix $L$, time complexity of this step is proportional to the number of $f_{mns}$ coefficients and, 
therefore, is equal to $O(N^3)$. This stage of the algorithm can be executed much faster if the matrix $L$ is sparse. 
Space complexity is $O(N^2)$ for both dense and sparse cases.

\textbf{2.5. Compute the coefficients $r_{sm}$ by formula (\ref{eq:09}).} This step calculates the matrix $R$ using the distribution of 
operations on the tensor $Z$ between MPI processes. Each process calculates groups of columns of the matrix $R$. 
To do this, it computes only nonzero terms $l_m \times f_{mns}$, $l_m \times z_{mns}$ and fills corresponding group of columns of $R$. Upon completion, 
all processes transfer data to the master process.

Tensors $F$ and $Z$ are filled in such a way that each of their two-dimensional plane sections 
contains from $O(N)$ ('sparse' section) to $O(N^2)$ ('dense' section) elements. 
In our prior work \cite{c12}, we noted that there exists $O(N)$ 'dense' sections containing $O(N^2)$ elements, and $O(N^2)$ 'sparse' sections containing $O(N)$ elements. 
Therefore, for every $r_{sn}$ tensor the number of nonzero coefficients $z_{jln}$, $f_{kls}$, $z_{kln}$, $f_{jls}$ varies from $O(N)$ to $O(N^2)$ that 
results in  maximal complexity of every $r_{sn}$ calculation equals $O(N^4)$. Hence, calculating the product of the number of elements and time complexity of calculating of each element, we can estimate overall time complexity as follows. For 'dense' $s$-indexes and 'dense' $n$-indexes total time complexity should be equal to $O(N^6)$. But due to the specific structure of $F$ and $Z$ tensors it is $O(N^5)$ operations only. If one of the indices $s$ and $n$ is sparse and the other is dense, time complexity can be also estimated as $O(N^5)$ thanks to the structure of $F$ and $Z$. If both indices are sparse, we need $O(N^5)$ operations. During this step, the matrix $R$ is stored as a set of red-black trees (each row is stored as a separate tree), and, therefore, adding each calculated coefficient to the tree requires $O(logN)$ operations which leads to the total time complexity of the step equal to $O(N^5 log N)$.

This stage is the most expensive in terms of memory. Straightforward implementation requires $O(N^3)$ space for intermediate data and up to $O(N^4)$ space for the matrix $R$ depending on sparsity of matrix $L$. To reduce the size of intermediate data we implemented a multistage procedure. This procedure slightly slows down the calculation, but reduces maximum memory consumption. 
The tensor $F$ can be divided into blocks by the third index, and at each moment only 2 such blocks can be stored in memory.
Using this fact, each process calculates its portion of the columns of the matrix $R$ gradually, block by block. 
As a result, a process computes its portion of the data, reducing memory consumption when storing its part of the tensor $f_{mns}$.

\textbf{2.6. Compute the initial value $v(0)$.} This step takes $O(N^2)$ time and $O(N^2)$ memory and can be done on every computational node.

\textbf{3. ODE integration (parallelized via MPI + OpenMP + SIMD).} 
During this step we integrate the linear real-valued ODE system (\ref{eq:04}) over time. While 
the Data Preparation Step is very memory consuming, this step is time consuming. Scalable parallelization of 
this step is a challenging problem because of multiple data dependencies. Fortunately, it does not take huge 
amount of memory and therefore can be run on smaller number of computational nodes than the Data Preparation Step. 
If the matrices $Q$ and $R$ are sparse, we employ the graph partitioning library ParMetis to minimize further MPI communications. 
Then we employ the forth-order Runge-Kutta method, one step of which takes $O(N^4)$ time for dense matrices $H$ and $L$ and $O(N^3)$ 
time for sparse matrices. The method requires $O(N^2)$ additional memory for storing intermediate results. 
Computations are parallelized via MPI on $K$ cluster nodes as follows. 
Matrices $Q$ and $R$ are split to $K$ groups of rows (panels) so that each portion of 
data stores approximately equal number of non-zero values. Then, each MPI process performs steps of the 
Runge-Kutta method for corresponding panels. The most computationally intensive linear algebra operations are 
performed by high-performance parallel vectorized BLAS routines from Intel MKL utilizing all computational cores and vector units.


\section{Algorithm performance}
\label{sec:5}

Numerical experiments were performed on the supercomputers Lobachevsky (University of Nizhni Novgorod), 
Lomonosov (Moscow State University) and MVS-10P (Joint Supercomputer Center of RAS). 
The performance results are collected on the following cluster nodes: $2 \times $Intel Xeon E5 2660 (8 cores, 2.2 GHz), 
64 GB of RAM. The code was built using the Intel Parallel Studio XE 17 software package.

The correctness of the results was verified by comparison with the results of simulations presented in the paper \cite{c12}. 
Later in this section we examine the time and memory costs when 
integrating sparse and dense models. Note that Data Preparation and ODE Integration are  separable steps. 
Therefore, when analyzing performance we consider them as consecutive  calculation stages.

\begin{figure}[H]
\centering
\includegraphics[width=0.9\textwidth]{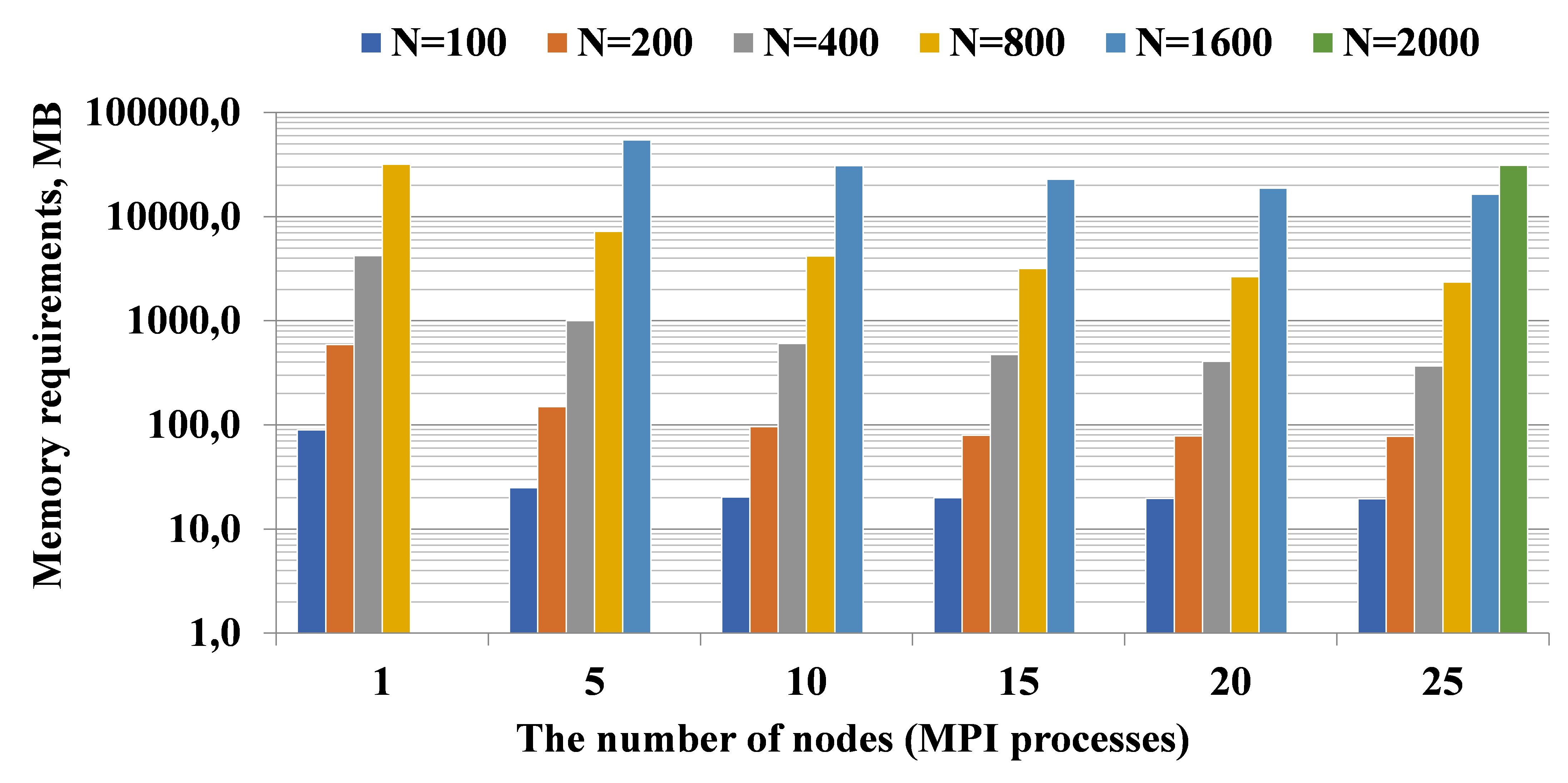}
\caption{Memory consumption per one node of the Data Preparation step for the sparse model.}
\label{fig:image2}
\end{figure}

\textbf{The Data Preparation step.} First, we consider the Data Preparation step and examine how increasing the dimension of 
the problem and the number of cluster nodes affect memory consumption. For the sparse model, we empirically found 
that it is advisable to perform 20 filtering stages when calculating the matrix $R$. Peak memory consumption when solving problems 
of size from $N=100$ to $N=2000$ is shown in fig. \ref{fig:image2}. 
Experiments show that when solving model of large dimension ($N=1600$) the memory requirements per node are reduced 
from 54.5 GB using 5 nodes to 16.5 GB using 25 nodes (scaling efficiency is $66.5\%$). Additionally, we were able to perform 
calculations for $N=2000$, which required 31 GB of memory at each of 25 nodes of the cluster. Computation time is 
significant but not critical for the Data Preparation step for the sparse problem. Nevertheless, we note that when using 
five cluster nodes computation time is reduced approximately by half compared to a single node and then decreases slightly.

\begin{figure}[H]
\centering
\includegraphics[width=0.9\textwidth]{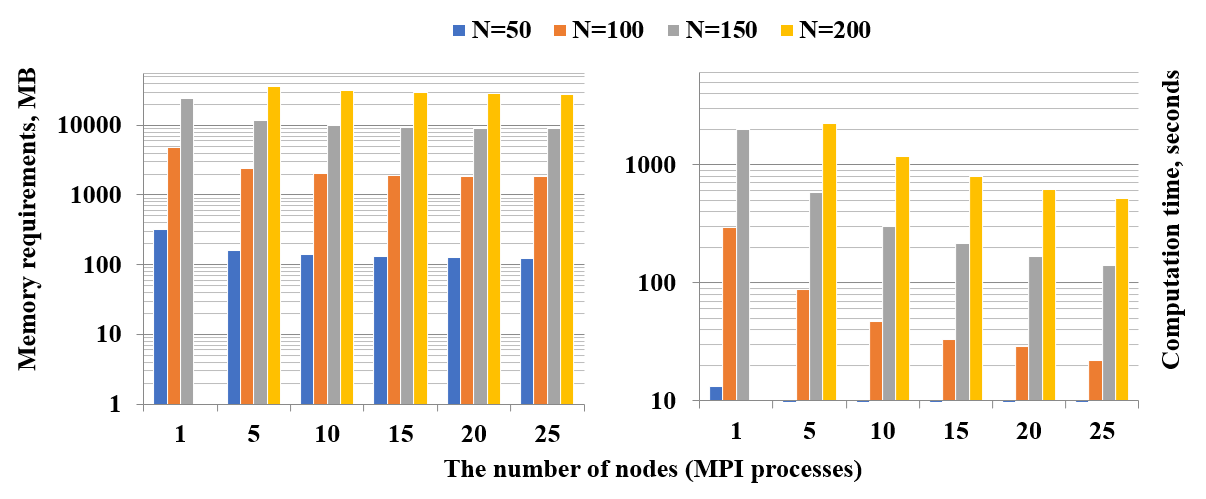}
\caption{Memory consumption per one node (left) and computation time (right) of the Data Preparation step for the dense model.}
\label{fig:image3}
\end{figure}

For the dense model, we managed to meet memory requirements on five nodes of the 
cluster upon transformation to the new basis of the problem of size $N=200$. 
On Figure~\ref{fig:image3} (left) we show how memory costs per node are reduced by increasing the number of 
nodes from 1 to 25. Note, unlike the case of the sparse model, the time spent on data 
preparation decreases almost linearly (fig. \ref{fig:image3}, right), which is an additional advantage of the parallelization.

\begin{figure}[H]
\centering
\includegraphics[width=0.9\textwidth]{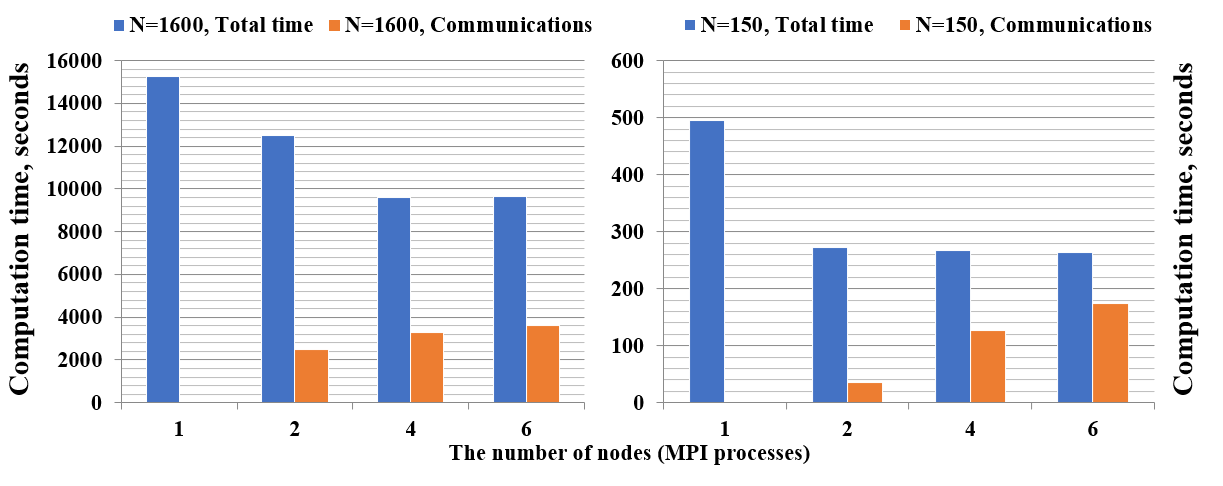}
\caption{Computation time of ODE Integration step for the sparse (left) and dense (right) models. 
Numerical integration was performed for one period of modulation $T$, with $20000$ steps per period.}
\label{fig:image4}
\end{figure}

\textbf{The ODE Integration step.} Next, we consider the ODE Integration step. 
In contrast to the Data Preparation step, where we concentrated on satisfying 
memory constraints, parallelization of this step is performed in order to reduce 
the computation time. Below we show the dependence of computation time and the number 
of cluster nodes used for the sparse (fig. \ref{fig:image4}, left) and dense (fig. \ref{fig:image4}, right) 
models. First, we load previously prepared data and run ODE Integration step for the sparse model of the size $N=1600$. 
The results show that it is enough to use only four nodes of the cluster. Further increase in the number of nodes does 
not reduce computation time due to MPI communications. When solving other sparse models, similar behavior is 
observed. Second, we load the precomputed data and run the ODE Integration step for the dense model of the size $N=150$. We 
found that the increase in the number of cluster nodes  quickly leads to an increase of the communications time, 
so it is enough to use two nodes to solve a dense problem of that size.

\section{Discussion}
\label{sec:6}

We presented a parallel version of the algorithm to 
model evolution of open quantum systems described with a master equation of the Gorini–Kossakowski–Sudarshan–Lindblad (GKSL) type.
The algorithm first transforms the equation into a system of real-valued ordinary differential eqautions (ODEs)
and then integrates the obtained ODE system forward in time. The parallelization is implemented for two 
key stages that are \textit{Data Preparation} [step (i)] for the transformation of 
the original GKSL equation into an ODE system and \textit{ODE Integration} [step (ii)] of the 
ODE system using the fourth-order Runge-Kutta scheme. The main purpose of the  parallelization of the first stage 
is to reduce memory consumption on a single node. We demonstrated that 
the achieved efficiency is enough to double the size of the sparse model compared to the sequential algorithm. 
In the case of the dense model, the run time of Data Preparation Step decreases linearly with 
the number of the nodes. Parallelization of ODE Integration step allows us to 
reduce the computation time for both, the sparse and the dense, models. Our implementation allows 
us to investigate the sparse model of the dimension $N=2000$ and the dense model of the dimension $N=200$ 
on a cluster consisting of $25$ nodes with $64$ GB RAM on each node.

\begin{figure}[H]
\centering
\includegraphics[width=0.8\textwidth]{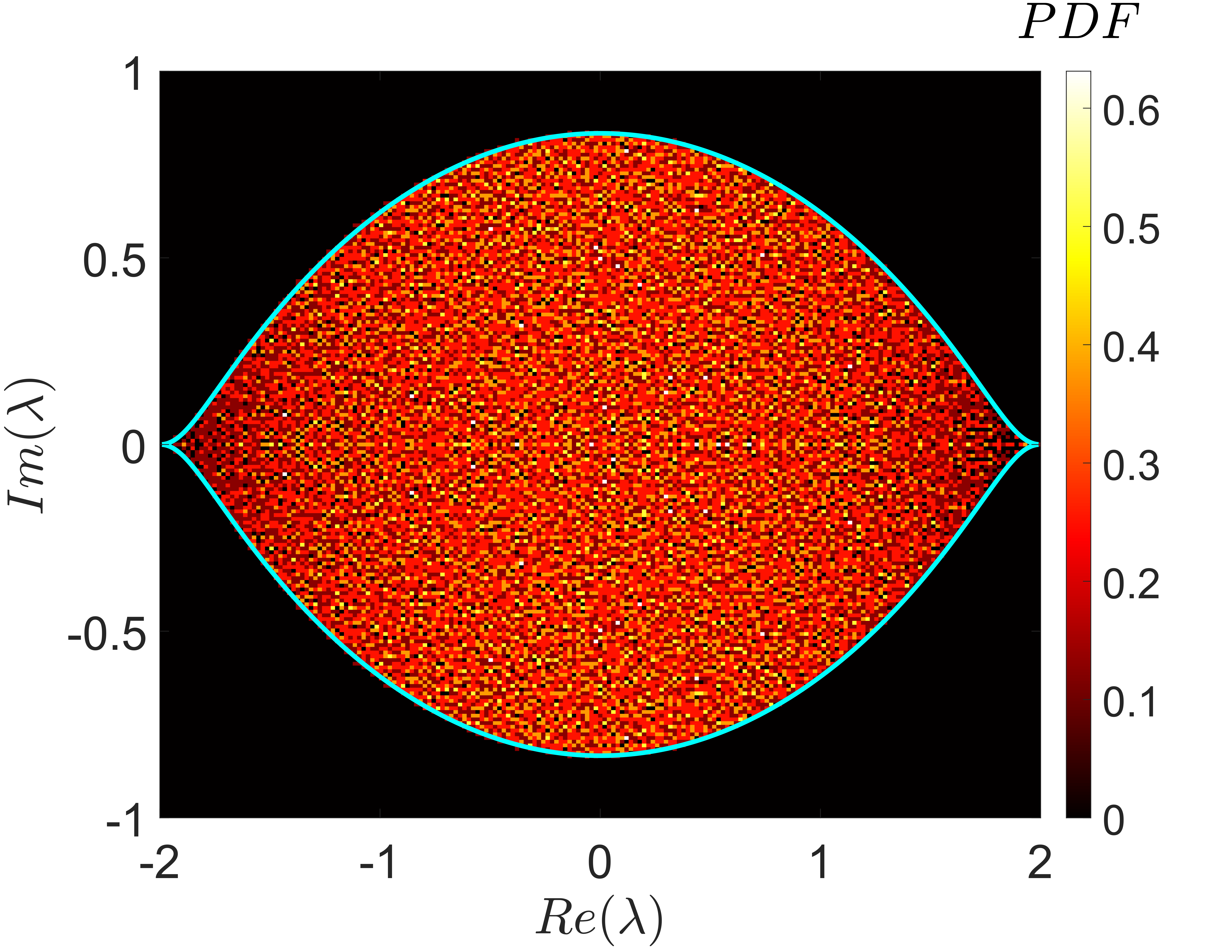}
\caption{Histogram of the complex eigenvalues, $\lambda_i$, $i = 2,3,....,N^2$, of a single realization of a 
random Lindbladian (see Section 3) for $N=200$. The shown area was resolved with a grid of $100 \times 100$ cells;
the number of eigenvalues in every cell was normalized by the cell area.
Altogether, $N^2 - 1 = 39 999$ eigenvalues are presented (except $\lambda_1 = 1$).
The bright thick line corresponds to the universal spectral boundary derived analytically in Ref.~\cite{randomL}.}
\label{fig:image5}
\end{figure}

The parallel version allows to explore spectral statistics 
of random Lindbladians acting in a Hilbert space of the dimension $N=200$; see Figure~\ref{fig:image5}.
As any statistical sampling, sampling over an ensemble of random Lindbladians is 
an embarrassingly parallel problem. However, the calculation of a \textit{single}
sample in the case of dense Lindbladian requires huge  memory costs when  $N > 100$.
Therefore, an efficient distribution of these costs among cluster nodes is needed. 
We overcame this problem by using  a two-level parallelization scheme. At the first level, 
we use trivial parallelization, in which each sample is calculated by a small group of nodes. 
At the second level, every group of nodes uses all available computing cores and memory to 
work with one sample. Although the speedup at the second level is not ideal, parallelization 
solves the main problem, allowing us to fit into the limitations on memory size. The proposed parallel 
algorithm opens up new perspectives to numerical studies of large open quantum models and allows to advance further into 
the territory of 'Dissipative Quantum Chaos' \cite{randomL,david,prosen1,prosen2}.

\acknowledgments{The work is supported by the Russian Science Foundation, grant No. 19-72-20086. 
Numerical experiments were performed on the supercomputers Lomonosov-2 (Moscow State University), 
Lobachevsky (University of Nizhni Novgorod), and MVS-10P (Joint Supercomputer Center of RAS).}





\end{document}